\documentclass[twocolumn]{aastex63}

\definecolor{lightgray}{gray}{0.90}
\usepackage{pifont}
\usepackage{color, colortbl}
\usepackage{comment}
\newcommand*\rot{\rotatebox{-75}}

\newcommand{\Msun}{$\mathrm{M_{\sun}}$}

\newcommand{\gaia}{\textit{Gaia}}
\newcommand{\LSST}{VRO}

\newcommand{\cmark}{ \ding{51}}%
\newcommand{\xmark}{ \ding{55}}%

\newcommand{\unsim}{\mathord{\sim}}
\newcommand\Ntot{35}
\newcommand\Namc{9}

\received{\today}
\accepted{TBD}
\submitjournal{AJ}

\shorttitle{ZTF outbursting AM CVn}
\shortauthors{Van Roestel et al.}
\graphicspath{{./}{figures/}}

\begin{document}

\title{A systematic search for outbursting AM CVn systems with the Zwicky Transient Facility}

\correspondingauthor{Jan van~Roestel}
\email{jvanroes@caltech.edu}

\author[0000-0002-2626-2872]{Jan van~Roestel}
\affiliation{Division of Physics, Mathematics, and Astronomy, California Institute of Technology, Pasadena, CA 91125, USA}

\author{Leah Creter}
\affiliation{Pasadena City College, 1570 E Colorado Blvd, Pasadena,  CA 91106, USA}
\affiliation{Division of Physics, Mathematics, and Astronomy, California Institute of Technology, Pasadena, CA 91125, USA}

\author[0000-0002-6540-1484]{Thomas Kupfer}
\affiliation{Department of Physics \& Astronomy, Texas Tech University, Box 41051, Lubbock TX 79409-1051, USA}

\author[0000-0003-4373-7777]{Paula Szkody}
\affiliation{Department of Astronomy, University of Washington, Seattle, WA 98195, USA}

\author[0000-0002-4544-0750]{Jim Fuller}
\affiliation{Division of Physics, Mathematics, and Astronomy, California Institute of Technology, Pasadena, CA 91125, USA}

\author[0000-0002-0948-4801]{Matthew J. Green}
\affiliation{Department of Astrophysics, Faculty of Exact Sciences, Tel-Aviv University, Ramat Aviv, Tel-Aviv 6139001, Israel }

\author{R. Michael Rich}
\affiliation{Department of Physics \& Astronomy, Univ. of California Los Angeles, PAB 430 Portola Plaza, Los Angeles, CA 90095-1547, USA}

\author{John Sepikas}
\affiliation{Pasadena City College, 1570 E Colorado Blvd, Pasadena,  CA 91106, USA}

\author[0000-0002-7226-836X]{Kevin Burdge}
\affiliation{Division of Physics, Mathematics, and Astronomy, California Institute of Technology, Pasadena, CA 91125, USA}

\author[0000-0002-4770-5388]{Ilaria Caiazzo}
\affiliation{Division of Physics, Mathematics, and Astronomy, California Institute of Technology, Pasadena, CA 91125, USA}

\author[0000-0001-7016-1692]{Przemek Mr{\'o}z}
\affiliation{Division of Physics, Mathematics, and Astronomy, California Institute of Technology, Pasadena, CA 91125, USA}

\author{Thomas A. Prince}
\affiliation{Division of Physics, Mathematics, and Astronomy, California Institute of Technology, Pasadena, CA 91125, USA}

\author[0000-0001-5060-8733]{Dmitry A. Duev}
\affiliation{Division of Physics, Mathematics, and Astronomy, California Institute of Technology, Pasadena, CA 91125, USA}

\author[0000-0002-3168-0139]{Matthew J. Graham}
\affiliation{Division of Physics, Mathematics, and Astronomy, California Institute of Technology, Pasadena, CA 91125, USA}

\author[0000-0003-4401-0430]{David L. Shupe}
\affiliation{IPAC, California Institute of Technology, 1200 E. California Blvd, Pasadena, CA 91125, USA}

\author[0000-0003-2451-5482]{Russ R. Laher}
\affiliation{IPAC, California Institute of Technology, 1200 E. California Blvd, Pasadena, CA 91125, USA}

\author[0000-0003-2242-0244]{Ashish~A.~Mahabal}
\affiliation{Division of Physics, Mathematics, and Astronomy, California Institute of Technology, Pasadena, CA 91125, USA}
\affiliation{Center for Data Driven Discovery, California Institute of Technology, Pasadena, CA 91125, USA}

\author[0000-0002-8532-9395]{Frank J. Masci}
\affiliation{IPAC, California Institute of Technology, 1200 E. California Blvd, Pasadena, CA 91125, USA}



\begin{abstract}
AM CVn systems are a rare type of accreting binary that consists of a white dwarf and a helium-rich, degenerate donor star. Using the Zwicky Transient Facility (ZTF), we searched for new AM CVn systems by focusing on blue, outbursting stars. We first selected outbursting stars using the ZTF alerts. We cross-matched the candidates with \gaia\ and Pan-STARRS catalogs. The initial selection of candidates based on the \gaia\ $BP$-$RP$ contains 1751 unknown objects. We used the Pan-STARRS $g$-$r$ and $r$-$i$ color in combination with the \gaia\ color to identify 59 high-priority candidates.
We obtained identification spectra of \Ntot\ sources, of which 18 are high priority candidates, and discovered 9 new AM CVn systems and one magnetic CV which shows only He-II lines. Using the outburst recurrence time, we estimate the orbital periods which are in the range of 29 to 50 minutes. We conclude that targeted followup of blue, outbursting sources is an efficient method to find new AM CVn systems, and we plan to followup all candidates we identified to systematically study the population of outbursting AM CVn systems. 
\end{abstract}



\section{Introduction}\label{sec:intro}


AM CVn-type systems are hydrogen-deficient and helium-rich accreting white dwarf binaries. They are part of the family of cataclysmic variables (CVs): white dwarfs that are accreting mass from a donor via Roche lobe overflow \citep{warner1995}. For AM CVn binaries, the donors are fully or partially degenerate and very compact. They evolve from close binaries; a white dwarf with a low-mass white dwarf \citep{paczynski1967, tutukov1979} or a helium star companion \citep{savonije1986, tutukov1989, yungelson2008}. These close binaries start accretion at orbital periods of $\unsim5-10$ minutes and evolve to longer periods (up to 65\,minutes) as mass is transferred to the white dwarf. A potential third channel involves cataclysmic variable with an evolved companion \citep[e.g.][]{breedt2012,carter2013a, thorstensen2002, podsiadlowski2003}, for a review see \citet{solheim2010} and \citet{toloza2019}. Although thousands of AM CVn systems are expected to be present in our Galaxy \citep{carter2013}, only $\unsim 60$ AM CVn systems are currently known due to their intrinsically low luminosity, see \citet{ramsay2018} for a recent compilation.

AM CVn systems are interesting for a number of reasons. Because of their compactness and short orbital periods, they are an excellent tool to study accretion physics under extreme conditions \citep[e.g.][]{coleman2018}.
Their short orbital periods also mean that the orbital evolution is influenced by gravitational wave radiation. Several hundred nearby AM CVn systems will be detectable by the \textit{LISA} satellite and are one of the most abundant types of persistent \textit{LISA} gravitational wave sources \citep{nelemans2004,nissanke2012,kremer2017,breivik2018,kupfer2018}. 

They are also potential progenitors of rare transient events. \citet{bildsten2007} suggests that, as a layer of helium builds up on the white dwarf, recurring He-shell flashes can occur which would look like helium novae. 
The mass of the He-shell becomes larger and the time between flashes longer as the systems evolve to longer orbital periods. This can result in a very energetic `final-flash' which can be dynamical and eject radioactive material from the white dwarf, dubbed a `.Iax' transient. 
In addition, the donors in AM CVn systems are the final remnants of stellar cores and have masses of 0.01--0.1\,\Msun. By measuring the chemical abundance of the accretion flow and/or the polluted white dwarf atmosphere, we can directly measure the composition of the core of a star \citep{nelemans2010}. 

Finally, one of the main questions regarding AM CVn systems is which of the three formation channels is most important: the white dwarf channel, the He-star channel, or the evolved CV channel \citep{toloza2019}. \citet{nelemans2010} showed that the observed CNO abundances can potentially be used to constrain the formation channel of AM CVn systems. AM CVn systems often show N lines, and O in rare cases, but C has never been detected in the visible \citep[e.g.][]{rui01,mor03,roe06,roe07,roe09,kup13,car14,car14a,kupfer2015,kup16}. This suggests that the systems mainly evolve through the WD-channel.
The entropy (and therefore mass and radius) are also predicted to be different for each of the formation channels. \citet{copperwheat2011}, \citet{green2018} and \citet{vanroestel2021} used rare eclipsing AM CVn systems to measure the donor mass and radius, which are higher than expected for systems formed through the WD and He-star channels. Finding more eclipsing AM CVn systems is crucial to resolve this inconsistency.

While AM CVn binaries are all accreting DB (helium atmosphere) white dwarfs with degenerate donors, their observational characteristics vary significantly \citep{nelemans2004}. Their appearance (both photometric and spectroscopic) depends strongly on the accretion rate, which is strongly correlated with the orbital period. The accretion rate determines the behavior of the accretion disk \citep{kotko2012,cannizzo2015} and also determines the white dwarf temperature \citep[e.g.][]{bildsten2006}. 

Very short period AM CVn systems ($P\lesssim10$ minutes) have high accretion rates and are `direct impact' accretors. In these systems, there is no accretion disk and the accretion stream directly impacts the white dwarf. They are detectable with X-rays, for example, HM\,Cnc and V407 Vul \citep{hab95,ram02,marsh2004, roe10}. At slightly longer periods ($10\lesssim P \lesssim 22$ minutes), the accretion flow forms an accretion disk. The accretion rate is high and the systems are in a constant `high state' \citep[e.g.][]{roe06a,kupfer2015,wevers2016,green2018a}. 
Intermediate period systems ($22\lesssim P \lesssim 45$ minutes) form an accretion disk, and behave similarly to hydrogen-rich CVs; they show outbursts and superoutbursts \citep{kotko2012} and show flickering in their lightcurves \citep[see][]{duffy2021}. As the orbital period increases, the outburst recurrence time increases exponentially \citep{levitan2015}, and the luminosity of the disk decreases as the orbital period increases \citep{nelemans2004}. 
In long-period systems ($P\gtrsim45$\,minutes), the accretion rate is low, outbursts are very rare (recurrence times of $\gtrsim1$ years), and the disk only contributes a tiny fraction of the overall luminosity.

The currently known sample of $\unsim 60$ has been built up over the years using various methods. Many AM CVn systems (including AM CVn itself) have been identified by their blue color and identification spectra. Most recently, \citet{carter2013} used SDSS and \textit{Galex} colors to perform a systematic spectroscopic survey of AM CVn systems. The second main method of finding AM CVn systems is by their outbursts. \citet{levitan2015} used PTF to find cataclysmic variables and identified AM CVn systems with followup spectroscopy. \citet{isogai2019} also focused on outbursting CVs, but instead used high-cadence photometry to measure the superhump period and classify a system as an AM CVn. Searching for short-period variability was also used to identify a new AM CVn system \citep[e.g.][]{kupfer2015,green2018a,burdge2020}. Despite all these efforts, Fig. 5 in \citet{ramsay2018} shows that we are still missing a significant number of nearby AM CVn systems.

The Zwicky Transient Facility \citep[ZTF,][]{bellm2019,graham2019} began to image the sky every night starting in 2019  to study the dynamical sky. Difference images are automatically generated and `alerts' are generated for any 5-sigma source in the difference images \citep{masci2019}. The alerts are used to identify extra-galactic transients, but are equally useful to identify outbursting stars like cataclysmic variables. \citet{szkody2020} identified 218 strong candidates in the first year of ZTF operations, and \citet{szkody2021} identified another 278 the second year.

In this paper, we present the method and first results of a targeted spectroscopic survey of blue, outbursting sources to find the missing AM CVn systems. We use ZTF to find outbursting sources and use their \gaia\ \citep{GaiaeDR3} and Pan-STARRS colors \citep{PS1} to select blue sources only (Section \ref{sec:targetselection}). We obtained identification spectra of \Ntot\ sources (Section \ref{sec:followup}) of which \Namc\ show helium lines and lack any hydrogen. We discuss the nature of 10 systems that do not show any sign of hydrogen in Section \ref{sec:individualsystems}, and present the results in Section \ref{sec:results}. We evaluate the approach used to find the new systems and discuss potential improvements in Section \ref{sec:discussion}. We end with a summary of the results.

\section{Target selection} \label{sec:targetselection}
AM CVn systems with orbital periods from 22--50 minutes show outbursts that are similar to those seen in many hydrogen-rich cataclysmic variables \citep{levitan2015,ramsay2018}. The accretion rate and the white dwarf temperature are similar for short period hydrogen-rich CVs and $P\gtrsim22$\,min AM CVn systems \citep[e.g.][]{toloza2019}.
The main difference is the donor, which is very cold ($T<2000$\,K) in AM CVn systems that show outbursts, while for hydrogen-rich CVs, the donor is a red dwarf or brown dwarf. 
This means that theoretically, the optical colors of AM CVn systems in quiescence are bluer compared to the majority of CVs which have a red dwarf donor (see for example \citealt{carter2013,carter2014}). Although CVs with brown dwarfs donors will also appear blue, the number of candidate AM CVn systems can potentially be significantly reduced by focusing on CVs with blue counterparts. 

To demonstrate the feasibility of this approach, we performed a pilot project where we selected CVs from \citet{szkody2020} and \citet{breedt2014}. We ranked them by their \gaia\ DR2 $BP$-$RP$ color and used the William Herschel Telescope (WHT) to take spectra of 7 blue, unclassified sources which were observable during at the time (Table \ref{tab:idspectra} and see \citealt{szkody2020}). Five show Balmer emission lines, typical for hydrogen-rich CVs, but two showed no signs of hydrogen and are new AM CVn systems (see Section \ref{sec:individualsystems}). 

\begin{figure}
    \centering
    \includegraphics{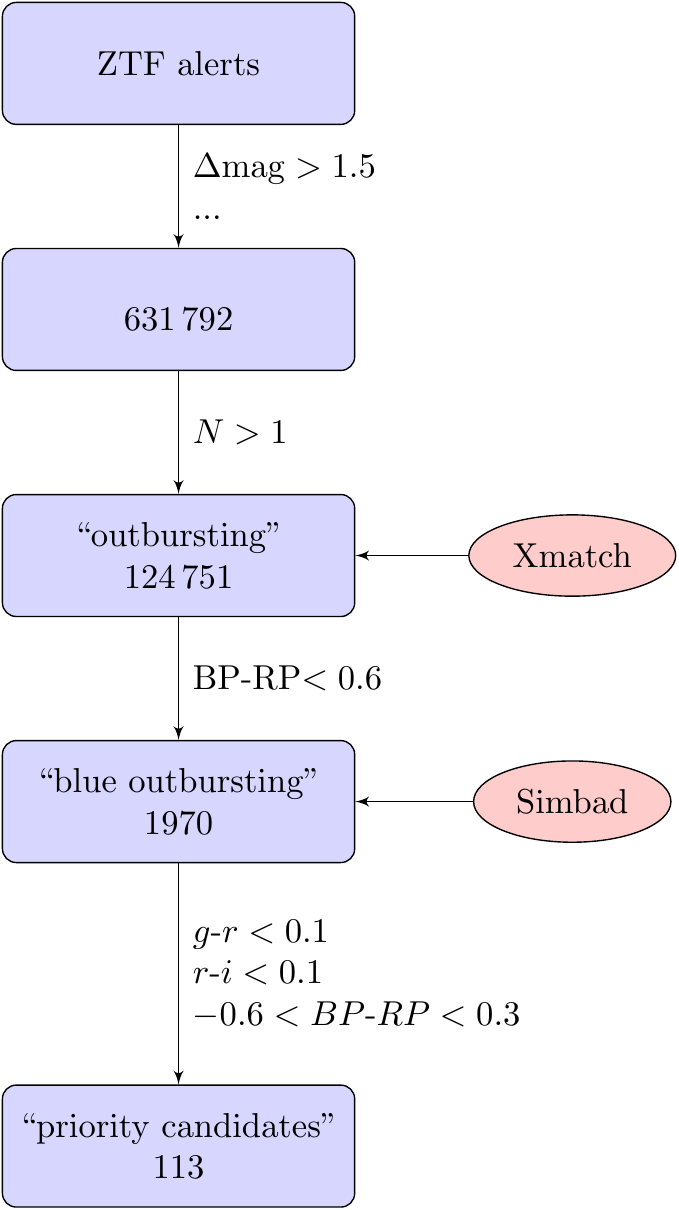}
    \caption{The different steps we used to select candidates. The number in each box shows the unique candidates that passed selection criteria so far. The selection criteria are explained in detail in the text, and a breakdown of the candidates in the last two steps are given in Table \ref{tab:candidates}.}
    \label{fig:flowchart}
\end{figure}

With the test a success, we performed a more systematic search. The different selection steps and the number of candidates are shown in Fig. \ref{fig:flowchart}. The next subsections discuss each step in detail.

\subsection{Selecting outbursting stars from ZTF}

As a first step, we used ZTF to identify outbursting stars. We searched all ZTF alerts \citep{masci2019} using a simple set of criteria to identify objects\footnote{\url{https://zwickytransientfacility.github.io/ztf-avro-alert/}}:
\begin{itemize}
    \item $\mathrm{isdiffpos}=\mathrm{True}$: only positive alerts.
    \item $\mathrm{distnr}<1.5$; alerts close to a star in the ZTF reference image.
    \item $\mathrm{psdistnr}<3.0$; alerts close to a star in PS1.
    \item $(\mathrm{magpsf}-\mathrm{magnr})<-2.5 \log_{10}(10^{0.4\times1.5}-1)$; select stars that become brighter by 1.5 mag or more.
    \item $\mathrm{braai}>0.9$ or $\mathrm{drb}>0.9$; reject bogus subtractions \citep{duev2019}.
    \item NOT $\mathrm{ssdistnr}>12$ \& $\mathrm{ssmagnr}<20$; used to remove known, bright asteroids.
    \item $\mathrm{magnr} > 16$; removes bright stars.
\end{itemize}
With these criteria, we selected 631\,792 unique objects.

After inspection of the candidates, we noticed that many objects with only a single alert, possibly bogus subtractions or asteroids that moved within 1'' of a star. We, therefore, removed any object for which only one alert passed the previous criteria:
\begin{itemize}
    \item $N>1$
\end{itemize}
after this step, 124\,751 ``outbursting'' objects remain.

\subsection{Crossmatch with Gaia and Pan-STARRS}
For the next step, we queried Vizier for both Gaia and PS1 data using Xmatch\footnote{\url{http://cdsxmatch.u-strasbg.fr/}}. Initially, we used Gaia DR2, but switched to Gaia eDR3 \citep{brown2020} when that became available. We only kept sources with: 
\begin{itemize}
    \item Gaia match within 3\arcsec
    \item $BP-RP<0.6$
\end{itemize}
These steps reduced the number of candidates to 1970.

\subsection{Identification of known sources}
\begin{figure*}
    \centering
    \includegraphics[width=\textwidth]{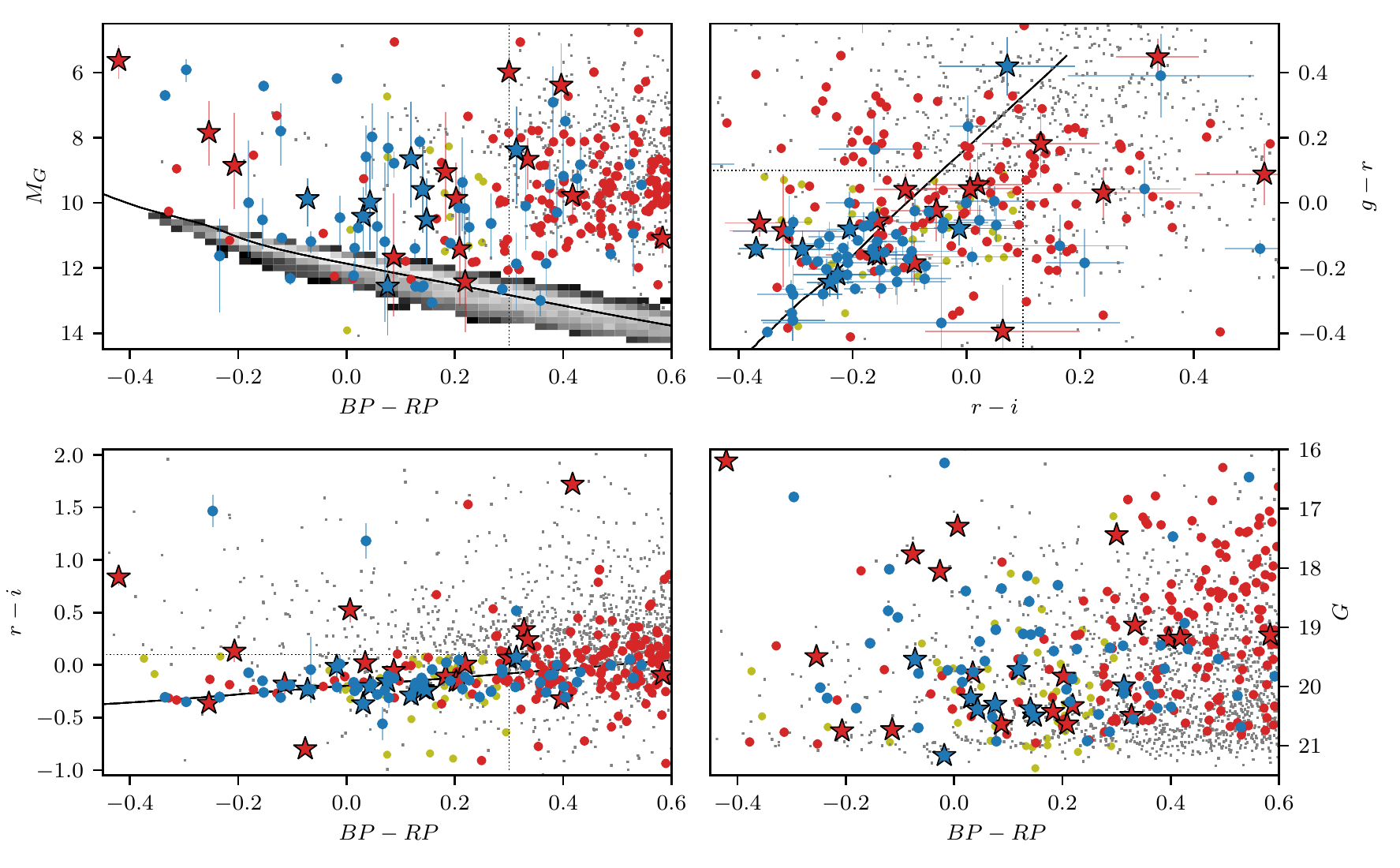}
    \caption{An overview of all blue outbursting sources found using our query. Blue markers show AM CVn and He-CV systems, red show other classified systems, yellow points show unidentified sources that passed all these criteria, the `priority candidates', grey points show other unclassified objects. The grey background in the topleft panel shows the population of nearby white dwarfs. Stars indicate ones for which we obtained an identification spectrum, see Table \ref{tab:idspectra}. Dashed lines show the `strict' selection criteria we used to prioritize targets for followup. The black lines show model colors of a DB white dwarf ($\log(g)=8$) \citep{bergeron2011}.}
    \label{fig:colorplots}
\end{figure*}

We listed all sources with their statistics that passed these criteria. We checked the literature on known AM CVn systems and identified 20 candidates as known AM CVn systems. To better understand where AM CVn systems are located in color space, we added all published AM CVn binaries to the list \citep{ramsay2018,green2020,vanroestel2021,isogai2021}, including 57 systems that did not appear in our alert query (they are not included in the numbers given in Fig. \ref{fig:flowchart} and Table \ref{tab:candidates}).
For the remaining candidates, we used Simbad\footnote{\url{http://simbad.u-strasbg.fr/simbad/}} to determine if a candidate was already identified as a hydrogen-rich CV. We searched for either an identification spectrum, a period from eclipses \citep{hardy2017}, or a period from superhumps  \citep{kato2017,patterson2005}, and marked these systems (199) as `not-AM CVn's and do not consider these any further. An overview of these sources in magnitude and color space is given in Fig. \ref{fig:colorplots}.

\subsection{Selection of priority candidates}
During the course of this work, Gaia eDR3 was released which includes more data and uses improved filter response curves. This changed the $BP$-$RP$ values for some objects, in some cases by as much as 0.6 mag, and changed the color of the false positives for which we already obtained spectra to much redder values.

The updated \gaia\ colors and an evaluation of the objects observed so far prompted us to use stricter color selection criteria. Inspection of the Pan-STARRS colors showed that many known AM CVn systems have blue colors in $g$-$r$ and $r$-$i$, while false positives are red in one or both of these colors (see Fig. \ref{fig:colorplots}). We, therefore, used the empirically chosen \citep[but see][]{carter2013} stricter set of criteria to identify high priority candidates:
\begin{itemize}
    \item $-0.6<BP$-$RP<0.3$
    \item $g$-$r<0.1$
    \item $r$-$i<0.1$
\end{itemize}

Table \ref{tab:candidates} summarises the properties of the selected candidates. This shows that the strict selection criteria reduce the number of candidates by an order of magnitude, but also excludes half of the known AM CVn systems. Based on the ratio of known AM CVn systems versus other systems, $\unsim$20\% of the objects that passed the strict criteria are AM CVn systems.
With 59 \textit{unidentified} systems that pass the strict criteria, we can expect to find $\unsim15$ new AM CVn systems by obtaining identification spectra of this sample.

\begin{table}
\caption{The details of the candidates selected using the initial selection criteria and the strict criteria. The `AM CVn' column shows the number of known AM CVn systems, the `H-CV' column shows the number of known hydrogen CV, the `unknown' column lists the number of unclassified objects. The rows indicate if \gaia\ measured a parallax regardless of uncertainty. }\label{tab:candidates}
\centering
\begin{tabular}{c|ccc|c}
 \multicolumn{5}{c}{\textbf{All candidates}} \\
& AM CVn & other CV & unknown & total  \\
\hline
\gaia\ plx. & 18 & 185 & 727 & 930\\
no plx. & 2 & 14 & 1024 & 1040 \\
\hline
total & 20 & 199 & 1751 & 1970 \vspace{0.5cm} \\ 
\multicolumn{5}{c}{\textbf{Strict selection}} \\
& AM CVn & other CV & unknown & total  \\
\hline
\gaia\ plx. & 9 & 39 & 24 & 72\\
no plx. & 1 & 5 & 35 & 41\\
\hline
total & 10 & 44 & 59 & 113 \\
\end{tabular}
\end{table}

\section{Followup observations}\label{sec:followup}

\subsection{Spectroscopic followup}
 We used the 4.2m William Herschel Telescope (La Palma, Sp) in June 2019 to observe 7 candidates with ACAM \citep[][]{benn2008}. ACAM has a resolution of $R\approx400$ and wavelength coverage of 4000--9000\AA.
We used exposure times of 900-1200 seconds. The spectra were reduced using the ACAM quick reduction pipeline. 

 We obtained 20 identification spectra with the 10m Keck\,I Telescope (HI, USA) and the Low Resolution Imaging Spectrometer (LRIS; \citealt{Oke1995,McCarthy1998}).
 Either the R600 grism for the blue arm ($R\approx1100$) and the R600 grating for the red arm ($R\approx1400$), or the R400 grism for the blue arm ($R\approx600$) and R600 grating for the red arm ($R\approx1000$) were used. The wavelength range is approximately 3200--8000/10000\AA.

 For two objects, we obtained spectra with DEIMOS \citep{faber2003} mounted on the 10m Keck II telescope. One spectrum was obtained with the 600ZD grating ($R\approx1400$), the second with the 1200B grating ($R\approx4500$). 
 Data were reduced with the standard pipeline.
 
We obtained spectra of 6 objects with the two-arm Kast spectrograph mounted at the 3-m Shane telescope \citep{miller1994}. In the blue and red arm, we used the 600/4310 and 600/5000 gratings. Combined with the 2\arcsec\ slit, the resolution is R$\approx$2200 and R$\approx$2500. We used exposure times between 1500 and 3600 seconds, depending on target brightness. We split the exposures in the red arm to mitigate the effects of cosmic rays. The data were reduced using a \textit{PipeIt} based pipeline \citep{prochaska2020}.

 \subsection{CHIMERA fast cadence photometry}
 We obtained $g$ and $r$ lightcurves of ZTF18acnnabo, ZTF18acujsfl, and ZTF19abdsnjm. We used CHIMERA \citep{harding2016}, a dual-channel photometer mounted on the Hale 200-inch (5.1 m) Telescope at Palomar Observatory (CA, USA). 
 Each of the images was bias subtracted and divided by twilight flat fields\footnote{\url{https://github.com/caltech-chimera/PyChimera}}.

 The ULTRACAM pipeline was utilized to obtain aperture photometry using a 1.5 FWHM-sized aperture \citep[see][]{dhillon2007}. 
 A differential lightcurve was created by simply dividing the counts of the target by the counts from the reference star. Images were timestamped using a GPS receiver.

\section{Individual systems}\label{sec:individualsystems}

\begin{figure*}
    \centering
    \includegraphics{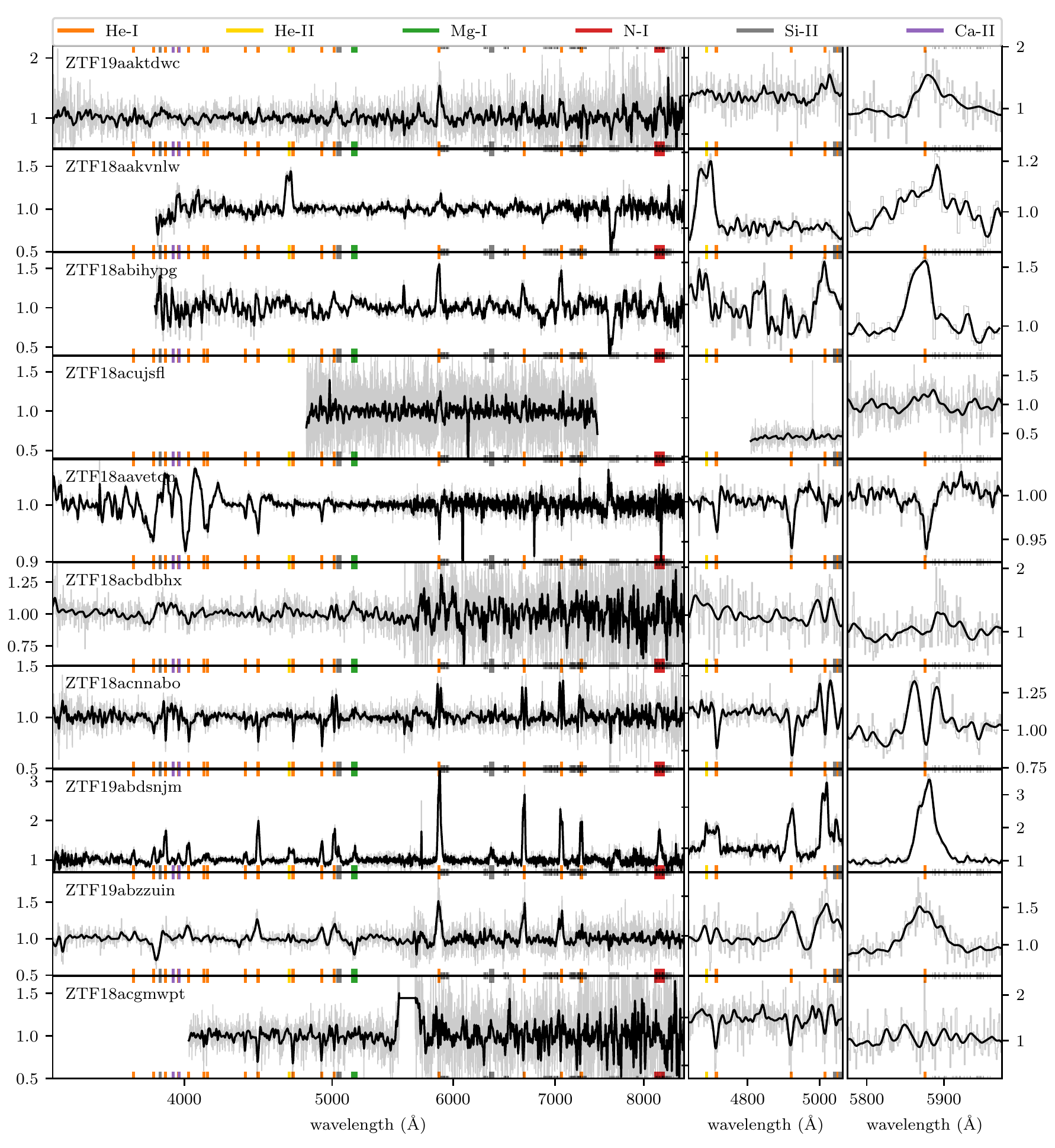}
    \caption{The spectra of the 10 systems which do not show any hydrogen in their spectra. The left panels show the entire spectrum and the two panels on the right zoom in on He spectral lines. The thin gray lines are original spectra and the thick black lines show spectra convolved with a Gaussian kernel for visibility. Vertical colored lines show the rest wavelength of different elements.}
    \label{fig:spectra}
\end{figure*}

\begin{figure*}
    \centering
    \includegraphics{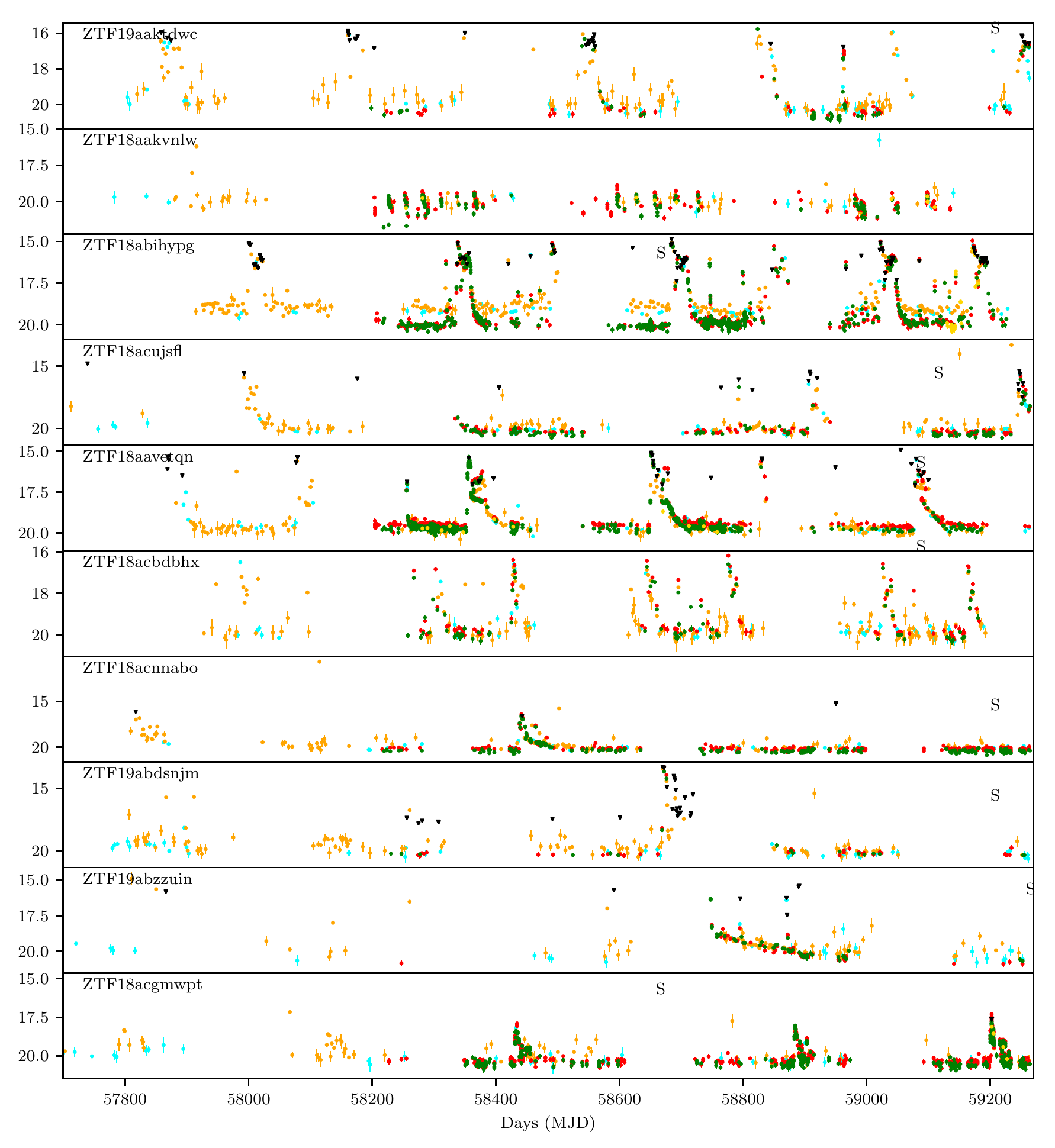}
    \caption{The lightcurves of the 10 systems which do not show any hydrogen in their spectra. We show ZTF forced photometry ($g$, $r$, and $i$ in green, red, and gold), ATLAS data ($c$ and $o$ in cyan and orange), and ASASSN data in black. An 'S' indicates when the spectrum was obtained.}
    \label{fig:lightcurves}
\end{figure*}

We obtained spectra of \Ntot\ objects of which 25 show hydrogen lines, shown in the Appendix.
Most spectra are typical for hydrogen-rich CVs with strong Balmer emission lines. A notable exception is the spectrum of ZTF19aadtlkv which shows a broad emission line, possibly cyclotron emission of a CV with a large magnetic field \citep[e.g][]{Szkody2003,szkody2020}.
The remaining 10 objects, shown in Fig. \ref{fig:spectra}, do not show hydrogen in their spectra. We discuss each of these objects individually in this section.

To characterize these objects, we obtained archival photometry for these sources, shown in Fig. \ref{fig:lightcurves}. The figure shows ZTF forced photometry \citep{masci2019}, ATLAS forced photometry \citep{tonry2018,smith2020}, and ASAS-SN \citep{shappee2014,kochanek2017} data for each object.

Similar to SU UMa-type cataclysmic variables, AM CVn systems show normal outbursts and also super-outbursts \citep{warner1995}.
Super-outbursts last a few weeks and are 2--5 magnitudes in amplitude, while normal outbursts are much shorter, typically less than one night and are lower in amplitude. 
\citet{levitan2015} and \citet{cannizzo2019} showed that the outburst frequency, amplitude, and superoutburst duration correlate with the orbital period of the AM CVn systems. We do note that the recent discovery of a very long outburst with a rather small amplitude of a long period AM CVn system complicates this simple picture \citep{riverasandoval2020}.

To estimate the orbital period we use the superoutburst recurrence time because the large amount of data available makes this easy and the correlation is the strongest. We visually inspected and marked the superoutburst peak times. We calculated the approximate large common divisor of the time differences. If there were multiple solutions, we inspected the lightcurve to determine which solution was best.

We convert the recurrence time to an orbital period using:
\begin{equation} \label{eq:1}
    P_\mathrm{orb} (\mathrm{min}) = 29.59 \left( \frac{\tau_\mathrm{rec}\mathrm(d) - 24.7}{100} \right) ^{0.136}
\end{equation}
which is the equation from \citet{levitan2015} rearranged. We assume a 5\% uncertainty or propagate the estimated variance of the recurrence time, whichever is larger.

\subsection{ZTF19aaktdwc}
This source was previously discovered by CRTS (CSS130419 J132918-121622, \citealt{drake2014}) as an unclassified outbursting star. A spectrum was obtained by \citet{Oliveira2020}, but it was completely featureless. The spectrum we obtained shows three weak, but significant, He-\textsc{I} emission line which confirms that this is an AM CVn binary.

The lightcurve shows at least 6 superoutbursts in the last 5 years that lasted a few weeks. There are also a few shorter and lower amplitude outbursts in between. The recurrence time of the superoutbursts is $\sim 210$d and very regular. This corresponds to an orbital period of $\unsim32.2$\, minutes.

\subsection{ZTF18aakvnlw}
This source was identified as an outbursting star by CRTS (CRTS J1647+4338) and also included in \citet{szkody2020}. Based on a spectrum which shows strong He-\textsc{II} emission and possibly H$\alpha$, \citet{breedt2014} speculate that this system is a He-CV (see also \citealt{green2020}). We also obtained a spectrum with ACAM, and also see a strong, double-peaked He-\textsc{II} line. A closer inspection of the spectrum suggests some He-\textsc{I} emission, but no sign of any H$\alpha$ emission. As already indicated by \citet{breedt2014}, strong He-\textsc{II} is typically associated with magnetic CVs. 

The lightcurve shows frequent, relatively low amplitude variability. The lightcurve does not show obvious superoutbursts and appears different from all other lightcurves. We cannot confidently classify this system as an AM CVn system or not. Phase-resolved spectroscopy is needed to determine the orbital period to definitively classify this object.

\subsection{ZTF18abihypg}
\citet{szkody2020} discovered this source as an outbursting star and a CV candidate. Despite being very blue and not passing the 'strict' criteria, we chose to get a spectrum of this source because the lightcurve showed many regular outbursts after the super-outbursts.

The spectrum shows clear, broad He-\textsc{I} emission lines. There is also a broad line at 8196\AA, likely N-\textsc{I}, sometimes seen in other AM CVn systems. Given the spectral features, we classify this source as a new AM CVn binary.

The lightcurve is similar to ZTF19aaktdwc with 6 (maybe 7) superoutbursts in the last 5 years with a recurrence time of $\unsim 160$d. This corresponds to a period of $\unsim 31.2$\, minutes. The frequency of regular outbursts is high in this system, with a total of 10 of them, typically lasting only a day.

\subsection{ZTF18acujsfl}
The spectrum is almost featureless but on closer inspection, three low amplitude, broad, and possibly double-peaked He-\textsc{I} emission lines can be seen. The very long outburst recurrence time ($\unsim 320$\,d) and low amplitude of the emission lines suggest that this is a longer period system, $P_\mathrm{orb}\unsim 34.3$\,minutes.

 The lightcurve shows just 2 (maybe 3) superoutbursts. There are possibly three regular outbursts, but these are not well sampled. We estimate the recurrence time to be $\unsim 310$\,d based on the interval between the last two observed outbursts. This time corresponds to an orbital period of $\sim 34.1$\, minutes. We obtained a 75-minute long lightcurve in $g$ and $r$ at a cadence of 5 seconds with CHIMERA. Both the $g$ and $r$ lightcurve did not show any sign of variability.

\subsection{ZTF18aavetqn}
This system was discovered by MASTER \citep{balanutsa2014}, detected by \gaia as Gaia18cjw, and was reported as a CV-candidate by \citet{szkody2020}. It is located in the Kepler field (KIC 8683556). The spectrum was obtained during the latest superoutburst and shows helium absorption lines, which are especially strong at the blue end of the spectrum. This is consistent with AM CVn systems in outburst.

The lightcurve shows 5 superoutbursts with 1 (maybe 3) normal outbursts. The time between superoutbursts is irregular, ranging from 170 to 290 days, with a best estimate of 230 days. This corresponds to a period of 32.6 (31.7--33.8) minutes. \citet{kato2017} reports a superhump period of $31.8$\,m which corresponds to an orbital period of $31.5$\,m (if we assume the same period excess of 0.8\% as was found for YZ LMi \citealt{copperwheat2011}). This is in good agreement with the estimated orbital period based on the superoutburst recurrence time.

\subsection{ZTF18acbdbhx}
\cite{drake2014} reported this as an outbursting star based on the CRTS data. The spectrum shows weak He-\textsc{I} emission lines. There seems to be a double-peaked line at He-\textsc{I} 5015\AA, and we, therefore, classify this as an AM CVn system.

The lightcurve shows 5--7 superoutbursts, with about 9 normal outbursts. The superoutburst recurrence time is regular at 130 days, which corresponds to a period of $29.4$\,m.

\subsection{ZTF18acnnabo}
This system was identified as an outbursting source by MASTER \citep{Shumkov2013}. The spectrum shows very clear double-peaked He-\textsc{I} emission lines in the red part of the spectrum. The He-\textsc{I} lines at the blue end of the spectrum appear as absorption lines. In addition, Ca-\textsc{I} H\&K lines are clearly visible in absorption. The very pronounced double-peaked structure of the He lines indicates that the inclination is high. This means that the system could potentially show eclipses. We obtained a 75-minute long lightcurve in $g$ and $r$ at a cadence of 5 seconds but there was no evidence of eclipses.

The lightcurve shows only two long outbursts, lasting at least 50 days followed by three normal outbursts. The outburst recurrence time is either 335 or 670 days. We used the MASTER detection and a PTF detection confirming the 670 day recurrence time. This puts the orbital period at $38.2$\,m.

\subsection{ZTF19abdsnjm}
This source is also known as ASASSN-19rg and Gaia19dlb. The spectrum is the most feature-rich of the entire sample. It shows very strong double-peaked He-\textsc{I} emission lines spanning the entire spectral range, as well as He-\textsc{II}-4686\AA. The spectrum also shows a Mg-\textsc{I} emission line at 5167/72/83\AA, and likely Si-\textsc{II} emission lines at 3856, 5175, 6347/71\AA. There is also a broad emission line at 8196\AA, likely N-\textsc{I} emission. 

The lightcurve shows only one superoutburst, with 5 shorter outbursts in the last 5 years. CRTS reports three detections on the night of MJD=56030, which puts an upper limit to the recurrence time to 7 years, corresponding to a period of 45\,m. 

 Josch H. and B. Monard\footnote{ \url{http://ooruri.kusastro.kyoto-u.ac.jp/mailarchive/vsnet-alert/23432}} observed the system while it was in outburst and found a superhump period of $43.8$\,m, consistent with our estimate. We obtained a 60-minute long lightcurve in $g$ and $r$ at a cadence of 5 seconds while the system was in quiescence. The system showed no sign of variability in the lightcurve.


\subsection{ZTF19abzzuin}
The spectrum shows He-\textsc{I} emission lines. The He-\textsc{I} lines at longer wavelengths show double-peaked profiles. The spectrum also shows a Mg-\textsc{I} absorption line at 5167/72/83\,\AA\ and Mg-\textsc{II}\ absorption line at 3832/38\,\AA.

The lightcurve shows only the tail end of one long superoutburst. The first detections could show the actual outburst, but the decay is fast enough to be a normal outburst similar to one seen later during the decay. The duration of the outburst and long recurrence time suggests a period $\gtrsim 40$--$50$\,m.

\subsection{ZTF18acgmwpt}
This source has been detected by \gaia\ as Gaia18djd. Because the SNR is low, the spectrum does not show any obvious features. A closer inspection shows He-\textsc{I} absorption lines on the blue side. The SNR is too low in the red to discern any features. We do classify this system as an AM CVn because of the detection of He-\textsc{I} lines.

The lightcurve shows three clear and two likely superoutbursts with a relatively low amplitude of 2 magnitudes. In the tail of each outburst, there are signs of normal outbursts. The outburst recurrence time is either 155 or 332 days, corresponding to orbital periods of $30.7$ or $34.5$ minutes.

\section{Results}\label{sec:results}

\begin{figure}
    \centering
    \includegraphics{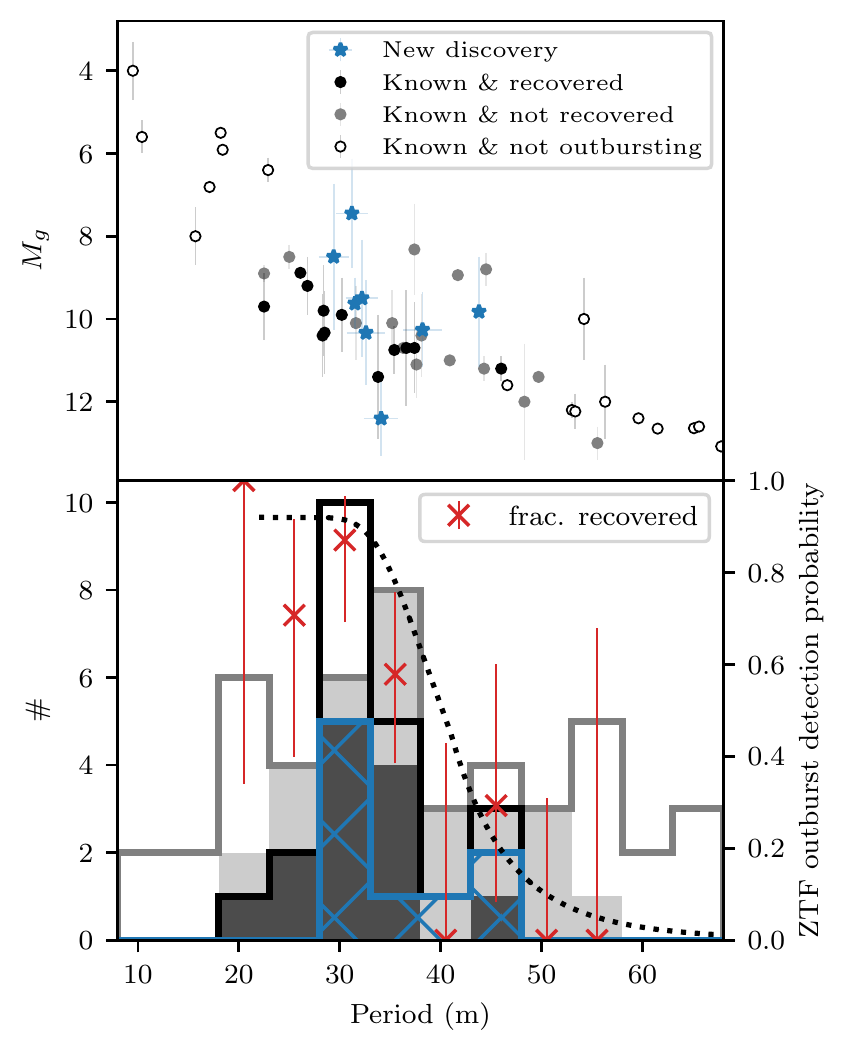}
    \caption{The top panel shows the absolute magnitude versus the orbital period for AM CVn for which both parameters are known. Blue stars indicate the new discoveries, black dots show outbursting systems we recovered, grey dots show outbursting systems we have not recovered, and open circles show non-outbursting systems. The bottom panel shows the period distribution of \textit{all} AM CVn systems with known periods. New discoveries are indicated with the hashed blue histogram, the grey histogram shows the known outbursting AM CVn systems. The recovered systems are shown in darker grey. The grey line shows all previously known AM CVn systems, including systems that do not show alerts. The thick black line shows all AM CVn systems recovered in our search. Red crosses show the recovery fraction of known systems and the dashed line shows the estimated probability that ZTF could have observed at least one outburst.}
    \label{fig:MG_P}
\end{figure}

\subsection{Selection of AM CVn candidates}
We identified 1970 blue, outbursting sources using ZTF, \gaia, and Pan-STARRS, of which 1751 are not classified. Based on colors of known AM CVn and other outbursting systems, we defined a simple and strict set of criteria designed to optimize the AM CVn discovery rate. A total of 113 sources pass the strict criteria, of which 59 are unclassified. We obtained \Ntot\ identification spectra in total, 18 of which are part of the set of strong AM CVn candidates. Analysis of the spectra shows that 19 systems show some kind of hydrogen emission and are typical cataclysmic variables. Out of the 10 sources which do not show hydrogen, 9 sources are new AM CVn systems, (of which 8 passed the strict criteria), and one source for which the classification is uncertain.

\subsection{Characterization of 9 new AM CVn systems}
The 9 new AM CVn systems are typical for outbursting AM CVn systems \citep{duffy2021}.and show both outbursts and super-outbursts. In some cases, there are hints of `dips' a few days after the peak of a super-outburst. In addition, normal outbursts typically last less than one night and tend to be more frequent after a super-outburst. We predict the orbital periods from the outburst recurrence time. The orbital periods are in the range of 29 to 34 minutes, with one system with a period estimated to be between 40--50 minutes. The distribution of periods shown in Fig. \ref{fig:MG_P}.

Spectroscopically the 9 new systems are diverse: some systems show strong emission lines, some show very weak emission lines, and other systems show absorption lines. Among the emission lines, some are double-peaked, while others show a single emission line. This diversity is also seen in known AM CVn systems and can be explained by differences in accretion state and inclination. A few systems, most notably ZTF18abdsnjm, also show metal lines including Ca, Mg, Si, and possibly N, also seen in other AM CVn systems \citep{nelemans2004}.

\section{Discussion}\label{sec:discussion}
The goal of this work is to determine if and how efficiently AM CVn systems can be identified by combining ZTF-alerts and color information. Here, we briefly discuss the completeness, efficiency, biases, and limitations of this method. 

\subsection{Selection by outbursts}
The completeness of our method to find \textit{all} AM CVn systems is limited by our search to objects which show outbursts of 1.5 magnitudes or more. While this selection method reduces the number of candidates by 4 orders of magnitude  (Fig. \ref{fig:flowchart}), it does mean that we will only find AM CVn systems that outburst; in the period range of $22\lesssim P_\mathrm{orb}\lesssim50$\,m. To find the shorter and longer period AM CVn systems that do not show outbursts, other methods are needed \citep[e.g.][]{burdge2020,vanroestel2021}. 

Superoutburst amplitudes of AM CVn systems typically range from $2$--$5$ magnitudes \citep{levitan2015}, which is larger than the 1.5 magnitude limit we chose. This means that the recovery efficiency of outbursting AM CVn systems is only limited by the ZTF coverage, sampling frequency, and time baseline. To better understand these aspects, we compare our sample with the known sample of AM CVn systems, as shown in Fig. \ref{fig:MG_P}. Using the already known systems, we calculate the recovery efficiency. We recovered most of the systems with periods between $22\lesssim P \lesssim 40$ min. The objects we did not recover in this period range were either at low declinations or were too bright (these were excluded from the recovery fraction efficiency). At periods of $P \gtrsim 40$ min, the recovery efficiency decreases sharply. Inspection of the lightcurves shows that these objects did not outburst when ZTF was watching.



\subsection{Selection by outbursts and colors}
As can be seen in Fig. \ref{fig:colorplots}, AM CVn systems are typically bluer compared to the overall population of outbursting CVs. Of the known AM CVn systems, 61 out of 67 (91\%) pass the first color cut ($BP$-$RP<0.6$), while reducing the number of candidates by 3 orders of magnitudes. Based on the known systems that pass this cut, we can expect $\unsim 9\%$ of the objects to be an AM CVn system (based on the known systems, see Table \ref{tab:candidates}). As explained in Section \ref{sec:targetselection}, this sample is still too large to followup with long-slit spectroscopy and we use a second more strict color cut to prioritize targets. 
From the known systems, we estimated a $\unsim$20\% (10/54) AM CVn-rate of the `strict' candidates (Section \ref{sec:targetselection}). As shown in Table \ref{tab:idspectra}, 8 out of 18 observed `strict' candidates are AM CVn systems (44\%). We did focus on the blue-est systems and systems with a parallax measurement. In addition, we also used the known superhump periods of two targets to prioritise them for followup. If we exclude these two systems, we still reach an efficiency of 38\%. 

The `strict' criteria were chosen to increase the true positive rate, however, the trade-off is low completeness. The 'strict' criteria reduce the number of known, outbursting AM CVn systems by half (Table \ref{tab:candidates}). Isolated white dwarfs hotter than 8000\,K, (less than the coolest white dwarf in an AM CVn system) should pass the strict criteria \citep[e.g.][]{bergeron2011}. However, an inspection of AM CVn color versus orbital period shows that there are many redder systems in the $BP$-$RP$, $g$-$r$, and $r$-$i$ colors in the range of 22-30 minutes. If we apply the `strict' criteria, we notice that almost all systems in the 22-30 minutes do not pass these criteria. By focusing the followup spectroscopy on the bluest sources, we have biased ourselves \textit{against} systems with periods shorter than $\unsim30$ minutes.

This can be explained by the more frequent outbursts for short period systems. Outbursts can `scramble' color measurements, especially for non-simultaneous observations like Pan-STARRS. If the object is observed in outburst in one band and in quiescence in the other, the magnitude difference does not represent the color of the system (e.g. the lower-left panel of Fig. \ref{fig:colorplots}). This is, again, especially relevant for short-period systems that show frequent outbursts. 


\subsection{Improvements}
Overall, the method we use is straightforward and simple and both the efficiency and completeness can be improved in a number of ways. First, we can expand the selection of outbursting sources by using CRTS \citep{breedt2012,drake2014}, PTF \citep{law2009,rau2009}, and \gaia\ alerts \citep[e.g.][]{campbell2015}. The longer time-baseline of these surveys should improve the detection efficiency in the period range of \textit{long} period systems; $40\lesssim P_\mathrm{orb}\lesssim50$\,m. We also note that most of the known AM CVn systems are in the Northern hemisphere which suggests that there are many undiscovered AM CVn systems in the Southern hemisphere. For example, \gaia\ alerts \citep{campbell2015} combined with the \gaia\ quiescence colors can be used to find them. 

Second, as noted in the discussion, using average (Pan-STARRS) colors can be scrambled by outbursts which can cause false-negatives. To resolve this, we can use the ZTF, ATLAS, and Pan-STARRS multi-color lightcurves to measure the color \textit{only} when the system is in quiescence. This would alleviate the problem that outbursts introduce noise in the color measurements and solve the bias against frequently outbursting, \textit{short}-period systems.

In addition, we have not used the temporal information from the lightcurves to attempt to identify AM CVn candidates. For example, AM CVn systems tend to show many normal outbursts after a superoutburst, which is only rarely seen in hydrogen-rich CVs. The amplitude, duration, and recurrence time of superoutbursts are also characteristic and could be used to distinguish AM CVn systems from hydrogen-rich CVs. In addition, if the sampling is very high, a measurement of a superhump period could be used to identify AM CVn systems. 

Based on the true positive rate and number of candidates, we estimate that there are another 20-30 outbursting AM CVn systems to be found with a \gaia\ parallax measurement in the ZTF footprint. We speculate that a similar number can be found in the Southern hemisphere which is relatively unexplored, which means that there is the potential to double the known number of AM CVn systems.

\section{Conclusions}\label{sec:summary}
We have described a spectroscopic survey aimed at finding new AM CVn systems by focusing on blue, outbursting sources identified by ZTF. We specified a set of strict criteria based on Pan-STARRS and \gaia\ colors. 103 candidates pass these criteria and based on the number of known sources, we estimated that $25\%$ of them are AM CVn systems. A detailed analysis showed that focusing on the bluest sources increased the purity of the sample, but also produced a bias against discovering short-period AM CVn systems. 

We obtained spectra of \Ntot\ candidates (18 from the strict selection) and identified \Namc\ new AM CVn systems. All spectra show helium lines, either in emission or absorption. A few systems also show metal absorption lines. From the recurrence frequency of super-outbursts, we estimate that their orbital periods range from 28 to 40 minutes. We encourage observers to obtain high cadence photometry when these systems outburst to confirm the orbital periods.

For future work, we will obtain identification for all unidentified objects that pass the strict criteria. We will aim to improve completeness and efficiency by including data from other surveys, e.g. CRTS \citep{drake2014}, \gaia-alerts \citep{wyrzykowski2012}, ATLAS \citep{tonry2018}, Skymapper \citep{wolf2018}, BlackGEM \citep{groot2019}. Finally, we will explore methods that use the multi-color lightcurves directly to improve the identification of candidates. The goal is to find all AM CVn systems with \gaia\ parallax measurements with the goal to study the population properties (space density, Galactic distribution, period distribution etc.) but also to find rare AM CVn systems; eclipsing systems, systems with large magnetic fields, or systems with unusual chemical abundances.

In addition, developing efficient selection methods will also be important for \LSST\ which will obtain 5-band lightcurves and find many faint candidates for which followup spectroscopy will be expensive.

\vspace{5mm}
\facilities{P48(ZTF), WHT:4.2m (ACAM), Keck2:10m (LRIS\&DEIMOS), P200:5.0m (CHIMERA), Shane:3.0m (Kast)
}


\software{penquins,
          astropy \& astroquery  \citep{astropycollaboration2013,astropycollaboration2018}, 
          Lpipe \citep{perley2019}, 
          PypeIt \citep{prochaska2020}}

\begin{acknowledgments}

This paper is dedicated in part to the memory of Dr. Bill Whitney who was above all a great teacher and educator. For over 68 years from Caltech to MIT to JPL and back to Caltech, Bill has guided and nurtured both students and faculty from across the country and world, with his love of physics, science, and education. He was one of the founders of the Caltech Summer Undergraduate Research Fellowship (SURF) in 1979 under the principle that undergraduates can productively contribute to research, which was considered very novel at the time. Since then, the SURF undergraduate model has been replicated by many leading universities benefiting many thousands of students and faculty alike. His deep recognition for the efficacy of multi-generational science education is reflected in this paper as contributing author and SURF student Leah Creter is the student of also contributing author Dr. John Sepikas who in turn was the student of Dr. Bill Whitney. Nothing would have made him more proud.

Based on observations obtained with the Samuel Oschin Telescope 48-inch 
at the Palomar Observatory as part of the Zwicky Transient Facility project. ZTF is supported by the National Science Foundation under Grant No. AST-1440341 and a collaboration including Caltech, IPAC, the Weizmann Institute for Science, the Oskar Klein Center at Stockholm University, the University of Maryland, the University of Washington, Deutsches Elektronen-Synchrotron and Humboldt University, Los Alamos National Laboratories, the TANGO Consortium of Taiwan, the University of Wisconsin at Milwaukee, and Lawrence Berkeley National Laboratories. Operations are conducted by COO, IPAC, and UW.

This research has made use of the SIMBAD database, operated at CDS, Strasbourg, France; Astroquery \citep{ginsburg2019}; Astropy, a community-developed core Python package for Astronomy \citep{astropycollaboration2018,astropycollaboration2013}. Some of the data presented herein were obtained at the W.M. Keck Observatory, which is operated as a scientific partnership among the California Institute of Technology, the University of California and the National Aeronautics and Space Administration. The Observatory was made possible by the generous financial support of the W.M. Keck Foundation. Based on observations made with the William Herchel Telescope (WHT) operated on the island of La Palma by the Isaac Newton Group in the Spanish Observatorio del Roque de los Muchachos of the Instituto de Astrofísica de Canarias. 

This work has made use of data from the European Space Agency (ESA) mission {\it Gaia} (\url{https://www.cosmos.esa.int/gaia}), processed by the {\it Gaia} Data Processing and Analysis Consortium (DPAC, \url{https://www.cosmos.esa.int/web/gaia/dpac/consortium}). Funding for the DPAC has been provided by national institutions, in particular, the institutions participating in the {\it Gaia} Multilateral Agreement.

The Pan-STARRS1 Surveys (PS1) have been made possible through contributions of the Institute for Astronomy, the University of Hawaii, the Pan-STARRS Project Office, the Max-Planck Society and its participating institutes, the Max Planck Institute for Astronomy, Heidelberg and the Max Planck Institute for Extraterrestrial Physics, Garching, The Johns Hopkins University, Durham University, the University of Edinburgh, Queen's University Belfast, the Harvard-Smithsonian Center for Astrophysics, the Las Cumbres Observatory Global Telescope Network Incorporated, the National Central University of Taiwan, the Space Telescope Science Institute, the National Aeronautics and Space Administration under Grant No. NNX08AR22G issued through the Planetary Science Division of the NASA Science Mission Directorate, the National Science Foundation under Grant No. AST-1238877, the University of Maryland, and Eotvos Lorand University (ELTE). 

\end{acknowledgments}

\bibliography{references,reference2}
\bibliographystyle{aasjournal}



\appendix

 \begin{longrotatetable}
 \begin{deluxetable*}{llllrlllllllll}
 \tablecaption{An overview of all objects for which we obtained a spectrum. The column 'strict' indicates if the sources passed the strict color selection, see section \ref{sec:targetselection}. Distances are taken from \citet{bailer-jones2021}.} \label{tab:idspectra}
 \tablewidth{700pt}
 \tabletypesize{\scriptsize}
 \tablehead{
 \colhead{ZTF ID} & 
 \colhead{RA} &
 \colhead{Dec} & 
 \colhead{$G$} & 
 \colhead{$BP$-$RP$} & 
 \colhead{dist. (pc)} & 
 \colhead{spectrum} &
 \colhead{date} & 
 \colhead{\rot{Strict}} & 
 \colhead{\rot{AM CVn}} &
 \colhead{Notes} \\
 } 
 \startdata
 ZTF17aacmiwg & 10 53 33.8 & \phantom{-}28 50 35.5 & 19.18 & 0.42 & $1890^{+700}_{-650}$  & ACAM & 2019-06-25 & \xmark & \xmark & $H\alpha$ emission, Balmer and He-\textsc{I} absorption \\
  ZTF18aabxxwr & 13 15 14.4 & \phantom{-}42 47 46.9 & 19.92 & 0.69 & $1590^{+410}_{-290}$ & ACAM & 2019-06-25 & \xmark & \xmark & strong Balmer and He-\textsc{I} emission \\ 
  \rowcolor{lightgray} ZTF19aaktdwc & 13 29 18.5 & -12 16 22.3 & 20.37 & 0.14 & $1960^{+1250}_{-1320}$ & ACAM & 2019-06-25 & \cmark & \cmark & no Balmer lines, He-\textsc{I} emission, maybe He-\textsc{II} emission \\
  ZTF19aadtlkv & 13 48 01.9 & -09 17 41.7 & 20.73 & -0.11 & & ACAM & 2019-06-25 & \xmark &  \xmark & strong Balmer emission, broad synchrotron beaming at 5543\AA \\
  ZTF18aakvnlw & 16 47 48.0 & \phantom{-}43 38 45.0 & 20.01 & -0.25 & $2480^{+860}_{-850}$ & ACAM & 2019-06-25 & \xmark &  \textbf{?} & strong He-\textsc{II} emission, He-CV \citep{green2020} \\
  ZTF18abcxayn & 19 19 15.7 & \phantom{-}35 08 50.3 & 18.96 & 0.33 & $1960^{+980}_{-410}$ & ACAM & 2019-06-25 & \xmark & \xmark& $H\alpha$ emission, Balmer and He-\textsc{I} absorption \\
  \rowcolor{lightgray} ZTF18abihypg & 21 28 22.1 & \phantom{-}63 25 57.3 & 19.99 & 0.31 & $780^{+190}_{-100}$ & ACAM & 2019-06-25 & \xmark & \cmark& HeI emission\\
  \hline
  ZTF18abwjzpq & 02 24 36.4 & \phantom{-}37 20 21.4 & 20.63 & 0.21 & $860^{+440}_{-300}$ & DEIMOS & 2020-09-21 & \cmark &  \xmark &  strong double peaked $H\alpha$, weak double peaked Balmer and He-\textsc{I} \\
 \rowcolor{lightgray} ZTF18acujsfl & 04 49 30.1 & \phantom{-}02 51 53.7 & 20.30 & 0.08 & $1170^{+490}_{-450}$ & DEIMOS & 2020-09-21 & \cmark &  \cmark & double peaked He-\textsc{I} \\
  \hline
  \rowcolor{lightgray} ZTF18aavetqn & 19 18 42.0 & \phantom{-}44 49 12.3 & 19.54 & -0.07 & $1020^{+320}_{-160}$ & LRIS & 2020-08-19 & \cmark & \cmark & Mg-\textsc{II}\\
  ZTF18abobpds & 20 15 35.3 & -14 16 44.4 & 20.49 & 0.32 & & LRIS & 2020-08-19 & \xmark&  \xmark& strong Balmer, weak He-\textsc{I} emission \\
  ZTF19aavhqxe & 20 42 58.8 & -00 33 54.0 &	17.29 & 0.00 & & LRIS & 2020-08-19 & \xmark&  \xmark& moderate Balmer emission, weak He-\textsc{I} \\
  \rowcolor{lightgray} ZTF18acbdbhx & 21 08 20.6 & -13 49 09.3 & 19.71 & 0.12 & $1560^{+850}_{-430}$ & LRIS & 2020-08-19 & \cmark& \cmark& almost featureless, shows very weak weak He-\textsc{I} \\
  \hline
  ZTF18abhqcfi & 00 13 01.1 & \phantom{-}51 13 58.8 & 20.12 & 0.69 &  & LRIS & 2020-12-17 & \xmark & \xmark& strong Balmer, weak He-I emission\\  
  ZTF18abtnfzy & 06 26 09.6 & \phantom{-}67 45 31.3 & 20.18 & 0.64 & $1690^{+710}_{-490}$ & LRIS & 2020-12-17 & \xmark&  \xmark& strong Balmer, weak He-\textsc{I} emission\\  
  \rowcolor{lightgray} ZTF18acnnabo & 08 20 47.6 & \phantom{-}68 04 24.0 & 20.19 & 0.03 & $2450^{+1460}_{-830}$ & LRIS & 2020-12-17 & \cmark & \cmark& double peaked He-I\\    
  ZTF19aazgilw & 10 44 50.1 & \phantom{-}23 24 30.9 & 19.83 & 0.20 & $420^{+150}_{-70}$ & LRIS & 2020-12-17 & \cmark&  \xmark& Balmer emission and broad absorption, weak He-\textsc{I} emission \\
  ZTF18aabxycb & 13 15 14.4 & \phantom{-}42 47 44.6 & 19.92 & 0.69 & $1590^{+410}_{-290}$ & LRIS & 2020-12-17 & \xmark& \xmark&  strong Balmer emission, weak He-I\\ 
  \rowcolor{lightgray} ZTF19abdsnjm & 13 25 58.1 & -14 52 26.0 & 20.38 & 0.04 & $1160^{+520}_{-370}$ & LRIS & 2020-12-17 & \cmark & \cmark & He-\textsc{I}/\textsc{II}, Mg\textsc{II}, Si\textsc{II}, and N\textsc{I} emission \\
\hline
  ZTF19aamfvdm & 05 53 15.7 & -26 48 46.9  & 20.32 & 0.21 & $1000^{+780}_{-310}$ & LRIS & 2021-02-12 & \cmark&  \xmark& strong Balmer emission, He-\textsc{I} emission \\
  ZTF19aadqrmz & 07 07 08.0 & -00 56 41.1  & 20.42 & 0.18 & $2160^{+2090}_{-1040}$ & LRIS & 2021-02-12 & \cmark&  \xmark& strong Balmer, weak He-\textsc{I} emission \\
  \rowcolor{lightgray} ZTF19abzzuin & 08 44 19.7 & \phantom{-}06 39 50.2  & 21.16 & -0.02 & & LRIS & 2021-02-12 &  \cmark & \cmark & He-\textsc{I} emission, Mg\textsc{II} absorption \\
  ZTF18aczeouv & 11 38 35.6 & \phantom{-}04 44 54.8  & 19.13 & 0.58 &  $520^{+150}_{-90}$ & LRIS & 2021-02-12 & \xmark &  \xmark & double peaked Balmer and strong, narrow absorption in H and He-\textsc{I} \\
\hline 
  \rowcolor{lightgray} ZTF18acgmwpt & 07 01 15.8 & \phantom{-}50 23 21.5 & 20.50 & 0.14 & $1640^{+1290}_{-670}$ & LRIS & 2021-02-15 & \cmark &  \cmark &  Mg-\textsc{II} \\
  ZTF18aaacggd & 07 44 00.5 & \phantom{-}41 55 03.5 & 20.64 & 0.09 & $610^{+300}_{-230}$ & LRIS & 2021-02-15 & \cmark &  \xmark & double peaked $H\alpha$, weak $H\beta$ \\
  ZTF18achzexf & 08 29 25.3 & -00 13 47.5 & 16.20 & -0.42 & $1370^{+170}_{-150}$ & LRIS & 2021-02-15 & \xmark &  \xmark & strong Balmer and Paschen emission, weak He-\textsc{I} emission \\
  ZTF18aawqkva & 15 44 28.1 & \phantom{-}33 57 24.3 & 18.06  & -0.03 & & LRIS & 2021-02-15 & \cmark&  \xmark & outburst spectrum. Weak $H\alpha$ emission, Balmer and He-\textsc{I} absorption \\
  ZTF18abtugvf & 18 26 34.7 & \phantom{-}22 56 10.5 & 20.74 & -0.21 & & LRIS & 2021-02-15 & \xmark&  \xmark & Balmer emission \\
  ZTF18aavoyoq & 18 14 39.7 & \phantom{-}50 18 34.9 & 17.76 & -0.08 & & LRIS & 2021-02-15 & \cmark&  \xmark & broad Balmer emission \\
    \hline
    ZTF18abcxayn & 19 19 15.7 & \phantom{-}35 08 50.1 & 18.96 & 0.33 & $1960^{+990}_{-400}$ & Kast & 2021-04-13 & \xmark &  ? & low SNR, no obvious lines \\
    ZTF18abmouep & 16 36 18.3 & -01 15 06.2 & 19.20 & 0.40 & $5060^{+3120}_{-1660}$ & Kast & 2021-04-18 & \xmark & \xmark & H$\alpha$ emission\\
    ZTF18acecect & 07 28 42.8 & \phantom{-}53 37 39.3 & 17.44 & 0.30 & $2470^{+870}_{-400}$ & Kast & 2021-04-29 & \cmark & \xmark & Balmer emission lines  \\
    ZTF19aahvgyg & 09 08 52.2 & \phantom{-}07 16 39.3 & 19.77 & 0.03 & $1920^{+1030}_{-740}$  & Kast & 2021-04-29 & \cmark & \xmark? & low SNR, maybe Balmer emission \\ 
    ZTF19acecitx & 06 21 50.3 & -24 43 47.2 & 19.47 & -0.25 & $3460^{+3000}_{-1460}$ & Kast & 2021-04-29 & \cmark & ? & low SNR \\
    ZTF21aapvkyy & 19 17 08.3 & \phantom{-}30 00 24.56 & 20.92 & 0.64 & $3040^{+1170}_{-1300}$ & Kast & 2021-04-29 & \xmark & \xmark  & strong  H$\alpha$ emission, He-I emission \\
 \enddata
 \end{deluxetable*}
 \end{longrotatetable}


\begin{figure*}
    \centering
    \includegraphics{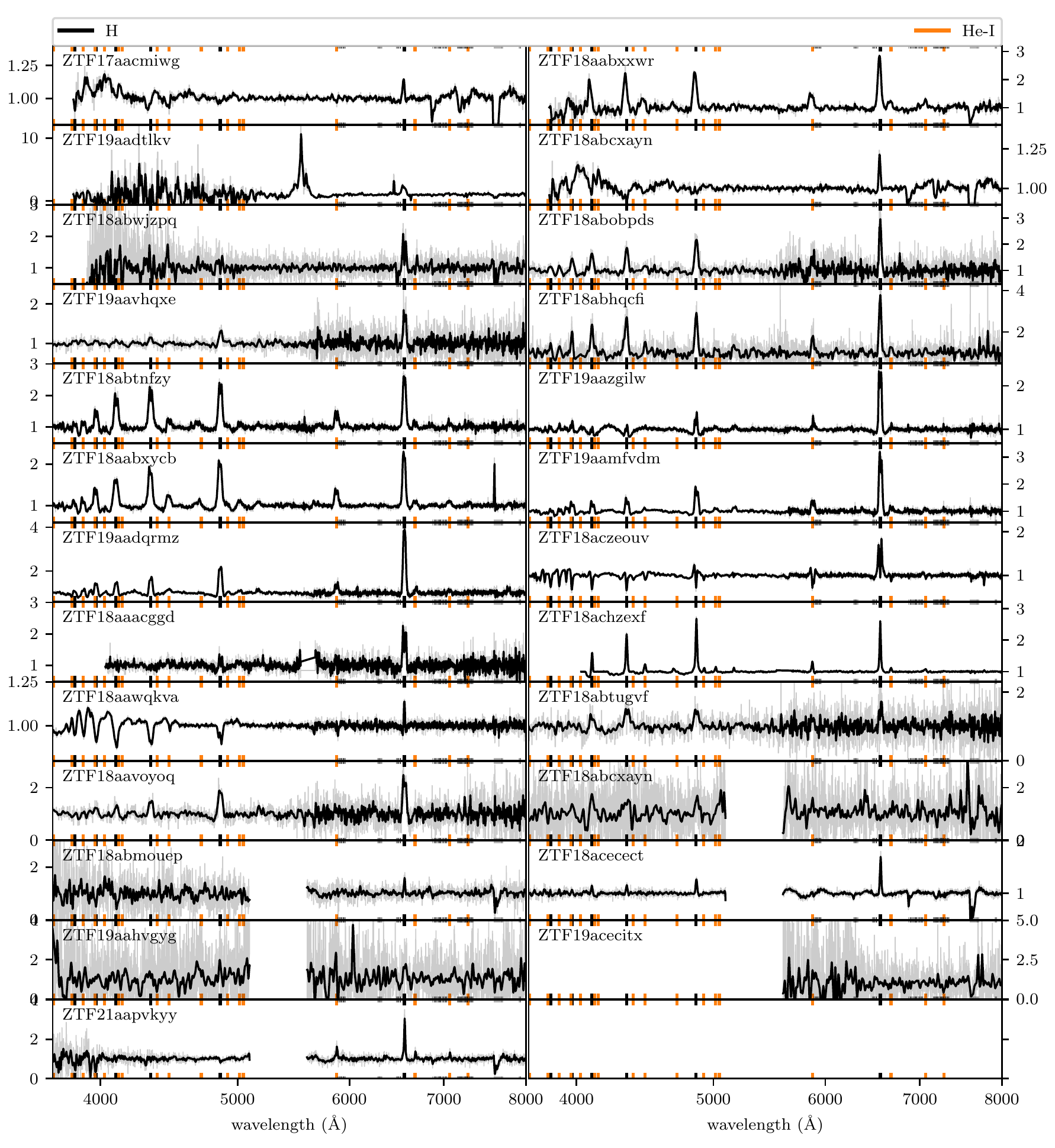}
    \caption{The spectra of that show hydrogen lines. The grey spectra show the original spectra and the black line show the spectra convolved with a Gaussian kernel for visibility. Vertical colored lines show the wavelengths of H and He lines.}
    \label{fig:otherspectra}
\end{figure*}

\end{document}